\documentclass[prx,aps,floatfix,longbibliography,superscriptaddress,twocolumn,final,notitlepage]{revtex4-2}
\RequirePackage{amsmath}
\RequirePackage{hyperref} 
\RequirePackage{amssymb}
\RequirePackage{graphicx}
\RequirePackage{bbm}
\RequirePackage{xcolor}
\usepackage{lipsum}
\usepackage{comment}

\usepackage{listings}

\lstdefinelanguage{json}{
    basicstyle=\ttfamily\small,
    showstringspaces=false,
    breaklines=true,
    frame=single,
    backgroundcolor=\color{gray!10},
    stringstyle=\color{black}, 
    keywordstyle=\color{black}, 
    commentstyle=\color{black}, 
    morestring=[b]",
    morekeywords={true,false,null} 
}

\usepackage{pgf}

\usepackage{mathtools}
\usepackage{caption}
\usepackage{subcaption}
\usepackage{amsthm}
\usepackage{graphicx}

\newtheorem*{definition*}{Definition}

\providecommand{\BibTeX}{\textsc{Bib}\TeX}

\usepackage{amsmath,fixmath}
\usepackage{setspace}

\usepackage{float}
\usepackage[nameinlink,capitalize]{cleveref} 

\usepackage{booktabs}
\usepackage{adjustbox}
\usepackage{tabu}
\usepackage[normalem]{ulem} 
\usepackage{soul}

\begin{document}

\title[
    Hypergraphx-data: a repository for higher-order network data
    ]{
    Hypergraphx-data: a repository for higher-order network data
}


\author{Quintino Francesco Lotito}
\email{lotitoq@ceu.edu}
\affiliation{Department of Network and Data Science, Central European University, 1100 Vienna, Austria}

\author{Lorenzo Betti}
\affiliation{Department of Network and Data Science, Central European University, 1100 Vienna, Austria}

\author{Berné Nortier}
\affiliation{Department of Network and Data Science, Central European University, 1100 Vienna, Austria}
\affiliation{School of Computer Science, University of St. Andrews, St. Andrews, KY16, Scotland}

\author{Alberto Montresor}
\affiliation{Department of Information Engineering and Computer Science, University of Trento, via Sommarive 9, 38123 Trento, Italy}

\author{Federico Battiston}
\email{battistonf@ceu.edu}
\affiliation{Department of Network and Data Science, Central European University, 1100 Vienna, Austria}
\affiliation{Department of AI, Data and Decision Sciences, Luiss University of Rome, Viale Romania 32, 00197, Rome, Italy}

\begin{abstract}
The availability of network datasets advances research in network science, machine learning and related fields by enabling empirical analyses and their reproducibility, algorithm development, model validation and benchmarking. Existing repositories, such as SNAP and Netzschleuder, have made traditional network datasets widely accessible with metadata, metrics, and basic visualizations. However, they primarily focus on pairwise interactions, limiting data access to systems with many-body interactions. To address this gap, we created \texttt{hypergraphx-data}, a repository of real-world hypergraph datasets for higher-order network analysis, spanning different domains from social networks to biology and finance, and supporting configurations such as weighted, directed, temporal, and multiplex hypergraphs. Each dataset includes relational information and metadata, provided in an open JSON format and a binarized format for Hypergraphx. We provide a user-friendly interface to facilitate browsing, filtering, and accessing the datasets, while also ensuring integrity and reproducibility through hash-based verification and data versioning. The repository is available at~\url{https://hgx-team.github.io/hypergraphx-data}.
\end{abstract}

\maketitle

\section*{Introduction}
The study of complex systems of interacting entities has gained widespread popularity over the last decades, establishing networks as a central tool in disciplines such as physics, biology, computer science, economics and the social sciences~\cite{newman2010networks}. A key driver of the growth of network science is the availability of datasets of empirical networked systems. These datasets are essential for developing and testing algorithms, validating theoretical models, exploring real-world patterns and phenomena, and enabling reproducibility. While some datasets are collected specifically for individual studies, 
the organized collection of publicly available datasets shared through data repositories has significantly broadened the spectrum of empirical analyses and the number of researchers able to test their theories and research ideas. Efforts such as the Stanford Network Analysis Project (SNAP)~\cite{snapdatasets}, the Network Repository~\cite{rossi2016interactive}, the Koblenz Network Collection (KONECT)~\cite{kunegis2013konect} and Netzschleuder~\cite{Netzschleuder} have played a major role in making network data accessible. These repositories provide datasets from various domains, along with metadata describing network properties, such as temporal activity of the edges or node attributes. Many of them also include filtering tools, precomputed metrics -- such as clustering coefficients and degree distributions -- and simple visualizations. 

Most existing repositories collect data representing traditional networks, where edges can only connect two system units. Over the last few years, the network science research community has shown that non-dyadic interactions are ubiquitous in many real-world systems~\cite{battiston2020networks}. Such higher-order interactions can be conveniently described by hypergraphs, where hyperedges encode interactions among an arbitrary number of system units~\cite{berge1973graphs}. Their presence has been shown to be crucial to understand the emerging dynamics and functions of many real-world complex systems~\cite{battiston2021physics}. Nevertheless, accessing large volumes of standardized and representative data of real-world systems with non-dyadic interactions is currently extremely challenging due to the fragmentation of data sources across domain-specific repositories, the common practice of collecting datasets solely for individual methodological studies without centralized maintenance, and ongoing technical limitations to integrate diverse data types and ensure data quality. All these issues hinder long-term accessibility, fair benchmarking of methods and effective data reuse.

In this work, we introduce \texttt{hypergraphx-data}, a repository of real-world hypergraph datasets, addressing the growing need for accessible, coherent, and heterogeneous higher-order data. The repository is associated with and complements Hypergraphx~\cite{lotito2023hypergraphx}, our open source library for higher-order network analysis, and aims to serve as a useful resource to facilitate and support research in higher-order network analysis, alongside other existing repositories, including XGI-data~\cite{landry2023xgi}, HypergraphRepository~\cite{antelmi2024hypergraphrepository} and \url{www.cs.cornell.edu/~arb/data/}. \texttt{hypergraphx-data} features a wide range of datasets from different domains, including social, biological and financial networks, and supports generalized hypergraph structures such as weighted, directed, temporal, and multiplex structures. Each dataset combines relational information and structural representations with rich metadata at the hypergraph, node, and edge levels, enabling both topological analysis and contextual insights. To ensure usability across different research contexts, datasets are provided in an open JSON format for interoperability, and a binarized format optimized for processing with the open-source library Hypergraphx (HGX)~\cite{lotito2023hypergraphx}. The repository also retains detailed information about data sources by crediting authors and providing clear citation guidelines to promote transparent and consistent data use. Finally, an intuitive website interface allows users to easily browse, filter and access the datasets. These contributions establish the repository as a useful resource for advancing the study of hypergraphs as modeling tools and for bridging their use toward real-world applications. At the time of publication, the repository contains $103$ datasets spanning $10$ domains. 

\section*{Data domains}
The repository provides data on both natural and artificial systems with higher-order interactions across multiple domains. Many of these systems have already been central in previous studies focusing on the limitations of traditional pairwise models, advocating for higher-order network modeling as a more appropriate framework for their analysis.

In social systems, higher-order interactions are common in activities involving groups of individuals. The repository includes datasets on human face-to-face interactions collected from settings such as schools, hospitals, and conferences~\cite{stehle2011high, vanhems2013estimating, mastrandrea2015contact, genois2015data, genois2018can}. Co-authorship datasets describe collaboration efforts in research, where groups of authors synergistically contribute to the same publication~\cite{sinha2015MAG, benson2018simplicial, pacs2021data}. Q\&A forum datasets capture online interactions where users collectively engage with the same questions, forming discussion groups~\cite{amburg2020hypergraph}. Voting data represent decision-making patterns involving groups of participants~\cite{justice2019database}.

In technological systems, group interactions appear, for example, in the context of communications and finance. In email datasets, hyperedges model interactions encoding messages sent to multiple recipients~\cite{leskovec2007graph, yin2017local, benson2018simplicial}. Bitcoin transaction datasets record financial exchanges involving multiple senders and receivers in a single transaction, providing insights into decentralized financial systems.

Ecological systems also exhibit higher-order interactions, such as in animal proximity networks. These datasets track interactions among groups of animals in their natural habitats, offering valuable data for studying collective behavior and ecological dynamics~\cite{gelardi2020measuring}.

In biology, higher-order interactions are often key to understanding complex systems. Gene-disease datasets connect groups of genes to specific conditions~\cite{bauer2011gene-disease}. Drug association datasets capture groups of drugs that interact or share similar properties~\cite{benson2018simplicial}. Metabolic reaction datasets represent groups of molecules involved in the same biochemical process, shedding light on metabolic pathways and their structure.

\begin{figure}
    \centering
    \includegraphics[scale=0.8]{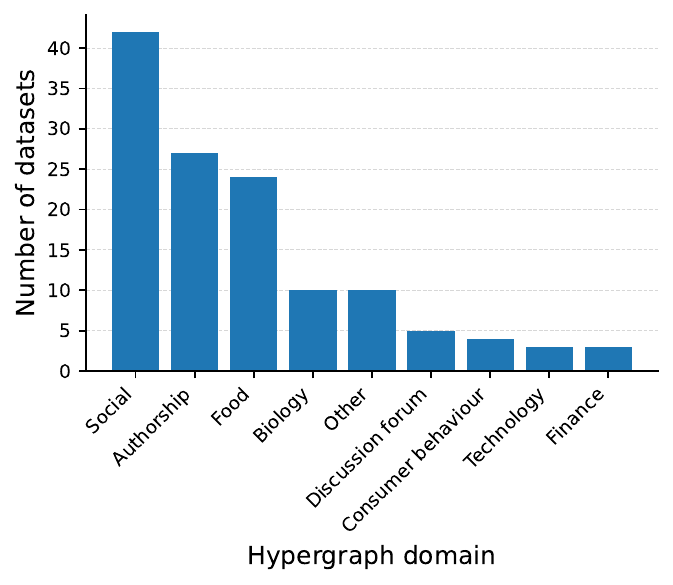}
    \caption{Distribution of datasets by application domain in the collection. Bars are sorted in descending order of frequency. We show only domains with at least three datasets.}
    \label{fig:data-stats-domains}
\end{figure}

In Fig.~\ref{fig:data-stats-domains}, we present statistics on the frequency of each domain relative to the total number of datasets.

\section*{Data features}
The repository hosts hypergraph datasets that encode both relational information, structural properties and metadata. In other words, each dataset provides not only the relationships among nodes encoded as a hypergraph structure,  but also the contextual information associated with the hypergraph as a whole, as well as its nodes and hyperedges.

The repository supports a wide range of hypergraph types. These include undirected hypergraphs, where simple hyperedges connect sets of nodes, suitable for contexts like group memberships or co-authorship networks. Directed hypergraphs extend the previous notion by specifying source and target nodes within hyperedges, making them ideal for modeling directional flows like information propagation or metabolic pathways. Weighted hypergraphs associate weights to hyperedges, quantifying the strength or importance of the relationships, useful in scenarios such as interaction frequencies in social networks. The repository also includes temporal hypergraphs, where hyperedges are associated with timestamps, capturing the evolution of relationships over time. These datasets are particularly relevant for studying dynamic systems, such as communication networks. Additionally, multiplex hypergraphs are supported, enabling multiple types of relationships between the same set of system units. In these cases, hyperedges are categorized into layers, each one representing a different type of interaction, such as distinct types of social ties.

In addition to structural features, the hypergraph dataset files include metadata, providing a richer set of contextual information at different levels. At the system level, hypergraph metadata describes the dataset as a whole, including its name, domain, version, source, and basic statistics about global properties such as average and maximum hyperedge size. At the unit level, node metadata captures attributes of the entities the nodes represent, such as age or sex in a social network, or molecular types in a biological network. 
At the interaction level, hyperedges are annotated with attributes such as weights, timestamps, or domain-specific labels, such as the field of a scientific paper or the genre of a movie. 

\begin{figure}
    \centering
    \includegraphics[scale=0.8]{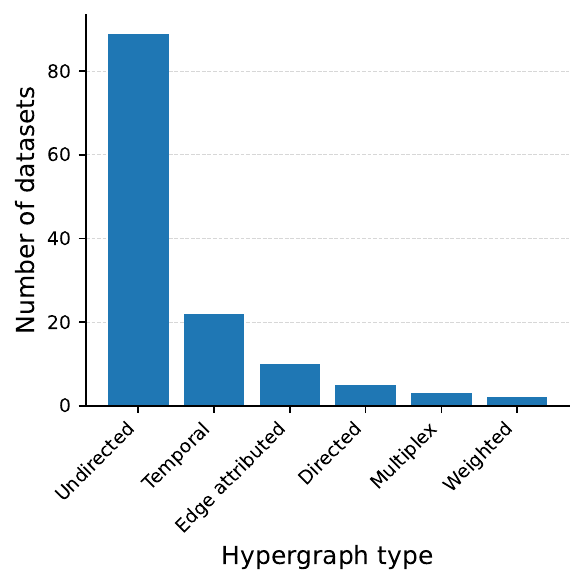}
    \caption{Distribution of datasets by hypergraph type in the collection. Bars are sorted in descending order of frequency.}
    \label{fig:data-stats-types}
\end{figure}

In Fig.~\ref{fig:data-stats-types}, we provide statistics on the distribution of hypergraph types as a fraction of the total datasets.

\section*{Data format}
The repository offers two dataset formats to balance accessibility and computational efficiency: a human-readable JSON format and a binary format optimized for loading with Hypergraphx~\cite{lotito2023hypergraphx}.

The JSON format is widely used and ideal for representing network data in a structured, self-descriptive manner. Nodes and hyperedges are stored as distinct objects. Nodes include attributes describing the entities (e.g., name or sex of a student), while hyperedges list their member nodes along with attributes such as weights, timestamps, or categorical labels. Global metadata provides information about the dataset's name, version, domain, and source. This format can be used by any library that supports our JSON schema by implementing a custom file reader.

To address performance requirements for large-scale datasets, the repository also provides a binary format. This format encodes hypergraph data as serialized Python arrays and dictionaries, minimizing storage overhead and significantly speeding up \texttt{I/O} operations, compared to JSON, which is on average 1.5$\times$ larger and about 5$\times$ slower~\footnote{Experimental setup: Ubuntu 24.04.3 LTS, x86\_64, 8 CPU cores, $\sim$94\,GiB RAM, Python 3.10; experiments were run single-threaded. We tested the entire dataset catalog, averaging loading times for ten runs. Storage sizes refer to uncompressed files.}.

Datasets are available for download in both formats and can be loaded into Hypergraphx using the \texttt{load\_hypergraph()} function, which reads files from a local path, or the \texttt{load\_hypergraph\_from\_server()} function, which automatically downloads and loads datasets directly from the remote repository. Moreover, Hypergraphx supports a communication channel compatible with the Hypergraph Interchange Format (HIF)~\cite{coll2025hif} through the \texttt{to\_hif()} function, enabling seamless data sharing with the other HIF-compatible libraries (such as XGI~\cite{landry2023xgi}, HyperNetX~\cite{praggastis2024hypernetx} and SimpleHypergraphs~\cite{spagnuolo2020analyzing}) without the need to implement custom schema readers. 

\section*{Data sources and processing}
The datasets collected in this repository come from three main sources.

We built several datasets from scratch by extracting raw information from publicly available websites and databases. For example, we constructed hypergraphs from IMDB data, where nodes represent actors and hyperedges represent co-participation in movies. Similarly, in ArXiv data, each paper is represented as a hyperedge connecting its co-authors. In transforming raw data into hypergraphs, we carefully preserve the original higher-order relationships without any loss of information.

We also include datasets that have already been introduced and distributed as hypergraphs by methodological papers across various fields such as network science, machine learning, and biology. These datasets are often scattered throughout the literature, so we have made a concerted effort to collect, centralize, and standardize them. This work ensures that our repository reflects the latest developments in hypergraph research and provides easy access to established benchmark datasets.

Another significant portion of the repository consists of datasets that were not originally distributed as hypergraphs, but whose structure is relevant for higher-order network analysis. In many cases, the original data are stored in forms such as bipartite incidence graphs, where one partition represents entities and the other groups, events, or relations. This is the case, for instance, of a variety of traditional data in ecology and biology. These datasets can be studied naturally within a hypergraph framework, making them suitable for inclusion in the repository. In some other cases, such as the face-to-face interaction data from the SocioPatterns project~\cite{sociopatterns}, data stored as simple graphs lose information about the original polyadic relationships. For these datasets, reconstructing a hypergraph representation requires additional information, such as fine-grained temporal information on each dyadic interaction, which can be exploited to heuristically infer group structure by identifying cliques of co-occurring dyadic ties, thus allowing these data to be analyzed as hypergraphs.

In order to ensure proper attribution and foster transparency, each dataset includes credits to the original work, with clear citation guidelines and easy access to \BibTeX code, licensing information and redistribution rights. We also link the exact source for raw data and store the scripts for building the derived hypergraph objects that we distribute in the repository. 

\section*{Data versioning and integrity}

The repository implements a system of semantic versioning and hash-based verification to effectively manage the evolution of datasets over time and support data integrity. This approach addresses an important challenge in the scientific community, since guaranteeing that shared datasets are accurately identified and remain consistent over time is fundamental for establishing transparent, fair and trustworthy benchmarks in research. 

Our versioning system is built on the simple principles of semantic versioning, which assigns a version number to each dataset in the format \textsc{major.minor.patch}.
This enables users to distinguish between significant changes, incremental updates, and minor corrections. We reserve major version changes for modifications that alter the structural properties of the networked systems, such as the addition or removal of new nodes or links. These changes may not maintain full compatibility with previous versions and may change the results of experimental analyses, for example, making clear identification essential. Minor updates reflect enhancements in the datasets, such as adding metadata to nodes and links, without altering the system structure. Patch updates address small corrections, such as fixing errors in node and link attributes. This versioning framework ensures transparency and allows users to consistently identify the exact dataset version and trace its evolution over time. It also guarantees that older versions continue to be available and recognizable, supporting reproducibility in research. To complement the versioning system, detailed changelogs are maintained for each dataset, documenting the nature of changes and the specific aspects of the dataset that were updated.

In addition to semantic versioning, the repository employs a hash-based system to verify dataset versions and ensure data integrity. Each version of a dataset is associated with a unique cryptographic hash (SHA-256), which serves as a digital fingerprint of the dataset. Release records provide immutable snapshots of the reference hashes and associated metadata at specific points in time, together with a changelog of additions, updates, and deprecated datasets. This approach allows users to independently verify the dataset they are working with by comparing the hash of their local file with the reference hash recorded by the repository for the corresponding version. A matching hash confirms that the dataset is unaltered and corresponds to the intended version. Any modification to the dataset results in a mismatched hash, signalling potential integrity issues.

\section*{Repository website}

\begin{figure*}
    \centering
    \fbox{\includegraphics[width=0.95\linewidth]{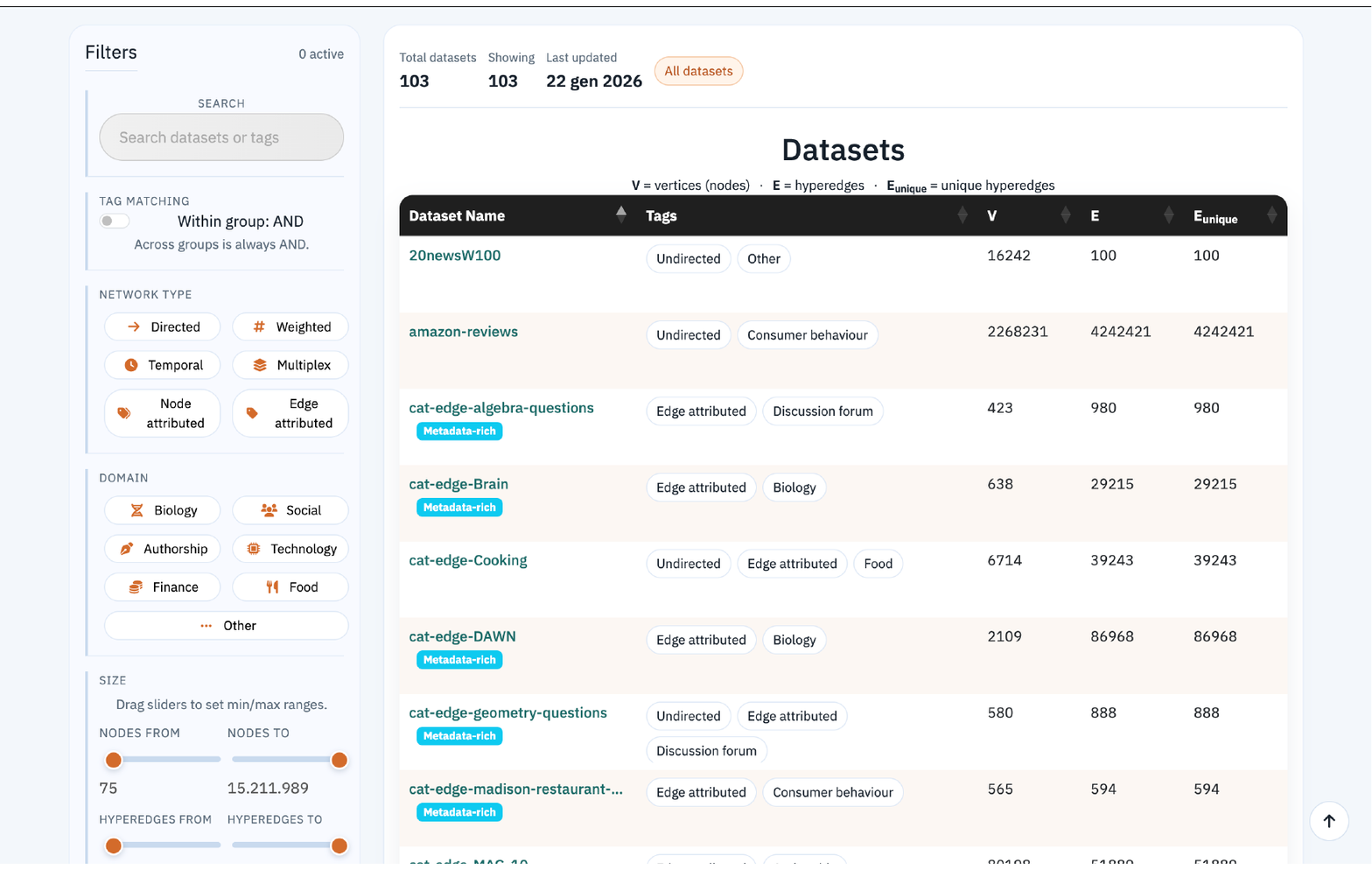}}

    \caption{Home page of the repository website, showcasing its interface and key features for navigating and accessing datasets.}
    \label{fig:screen-website}
\end{figure*}

The repository website is designed to be an intuitive interface for users to explore and download hypergraph datasets. The homepage provides an overview of the purpose and scope of the repository, with quick access to datasets and updates, such as recently added data or version changes. Users can navigate the dataset browser to explore the full data catalogue. The catalogue includes search and filtering tools, allowing users to narrow their options based on specific criteria. Datasets can be filtered by domain, such as social, biological, or technological networks, or by dataset characteristics, such as the number of nodes and the number of hyperedges, or by structural features including the presence of temporal annotations or weighted hyperedges. In Fig.~\ref{fig:screen-website}, we present the homepage of the website associated with our data repository. 

Each dataset is associated with a dedicated page that provides more details about the system, including a comprehensive overview of its metadata and characteristics. On these pages, users can preview key attributes of the selected dataset, such as its structural summary statistics, before downloading it. Moreover, such pages also contain information about current and older versions and supported formats, as well as appropriate source references in \textsc{BibTex} format. To offer a quick and informative snapshot of the dataset, we also provide basic data visualizations, such as hyperedge size and hyperdegree distributions, directly on the detail pages.

\section*{Conclusions}
\label{sec:conclusions}
Advances in data-driven research depend critically on robust software tools and the availability of open, reliable data repositories. In this work, we have introduced a centralized repository designed specifically for hypergraphs, addressing the increasing demand for an organized, diverse collection of datasets that capture complex systems with higher-order relationships. Our framework supports advanced hypergraph features such as temporal and multiplex structures, integrating relational information with system, node and interaction metadata. Complementing existing data sources, our repository aims to fill a critical gap in the community, giving network scientists access to a large, open and standardized amount of empirical data, enabling the test of new ideas and aiming to advance the development of network science beyond traditional dyadic approaches.

Besides the continuous addition of new datasets, future expansions of \texttt{hypergraphx-data} could include tools for interactive visualization of hypergraphs and expand their associated statistics, including temporal activity~\cite{cencetti2021temporal}, directed~\cite{lotito2026microscale} and multiplex measures~\cite{lotito2024multiplex}. We aim to design a more robust binary format, with an emphasis on safe loading and faster \texttt{I/O}, to better support large-scale datasets. We also aim to enhance the overall compatibility of the dataset collection with the Hypergraph Interchange Format (HIF) by contributing to extend the range of hypergraph types supported by the standard. Finally, we believe that this collection of datasets can serve as a foundation for establishing common benchmarks for performance-oriented tasks, such as motif mining~\cite{lotito2024exact}, community detection~\cite{contisciani2022inference, ruggeri2023community}, dimensionality reduction~\cite{kirkley2025structural} including filtering~\cite{musciotto2021detecting}, influence maximization~\cite{genetti2024influence}, or shortest paths computation~\cite{nortier2025higher}, similarly to the Open Graph Benchmark (OGB), which has become a standard for machine learning on graph problems.

\section*{Acknowledgements} 
F.B. acknowledges support from the Austrian Science Fund (FWF) through project 10.55776/PAT1052824 and project 10.55776/PAT1652425. A.M.
acknowledges support from the European Union through Horizon Europe CLOUDSTARS project (101086248).

\bibliographystyle{ScienceAdvances}
\bibliography{bibliography}

\end{document}